\documentclass[proof]{WileyASNA-v1}

\articletype{Article Type}%

\received{2022}
\revised{2022}
\accepted{2022}

\usepackage{threeparttable}
\usepackage{comment}
\usepackage{xspace}
\usepackage{hyperref}
\usepackage{url}
\usepackage{afterpage,array,rotating}
\usepackage{arydshln}
\usepackage{multirow}

\makeatletter
\newcommand{\figcaption}[1]{\def\@captype{figure}\caption{#1}}
\newcommand{\tblcaption}[1]{\def\@captype{table}\caption{#1}}
\makeatother

\newcolumntype{Y}{>{\centering\arraybackslash}p{4.5em}}
\newcommand{\vout}{\ensuremath{v_\mathrm{out}}\xspace}
\newcommand{\nh}{\ensuremath{N_\mathrm{H}}\xspace}
\newcommand{\xmm}{\textit{XMM-Newton}\xspace}
\newcommand{\iras}{IRAS~13224$-$3809\xspace}

\newcommand{\pl}{{\tt powerlaw}\xspace}

\newcommand{\zxipcf}{{\tt zxipcf}\xspace}

\begin{document}

\title{A novel  ``spectral-ratio model fitting'' to resolve  complicated  AGN X-ray spectral variations}

\author[1,2]{Takuya Midooka*}
\author[3,4]{Misaki Mizumoto}
\author[1,2]{Ken Ebisawa}

\authormark{Midooka \textsc{et al}}

\address[1]{Institute of Space and Astronautical Science (ISAS), Japan Aerospace Exploration Agency (JAXA), Kanagawa, Japan}
\address[2]{Department of Astronomy, Graduate School of Science, The University of Tokyo, Tokyo, Japan}
\address[3]{Hakubi Center, Kyoto University, Kyoto, Japan}
\address[4]{Department of Astronomy, Graduate School of Science, Kyoto University, Kyoto, Japan}

\corres{*~\email{midooka@ac.jaxa.jp}}

\abstract{Contemporary radiation-magnetohydrodynamic simulation of the AGNs predicts presence of the hot and strong accretion disk wind, which gets unstable far from the central region and turns into gas clumps.   
These inner-wind and outer clumps may be actually observed as the ultrafast outflows (UFOs) and the clumpy absorbers, respectively. We may call this picture as the ``hot inner and clumpy outer wind model''.  Observationally,  it is challenging to place  constraints  on the origin of the UFOs and clumpy absorbers due to complicated  spectral variations.
 We developed a novel method, ``spectral-ratio model fitting'', to resolve parameter degeneracy of the clumpy absorbers and other spectral components.  
 In this method, the parameters of the absorber in the line of sight are estimated from the ratio of the partially absorbed spectrum to the non-absorbed one.
We applied this method to the narrow-line Seyfert 1 galaxy \iras  observed by \xmm in 2016 for 1.5~Ms, where the source showed extreme spectral variability and complex absorption features. 
 As a result, we found that the soft spectral variation is mostly explained by a change of the covering fraction of the mildly-ionized clumpy absorbers, and that these absorbers are outflowing with such a high velocity that is comparable to that of the UFO ($\sim$~0.2--0.3 c).  
 This result implies that the formation of the clumpy absorbers and the UFO shares the same origin, supporting the ``hot inner and clumpy outer wind model''.
}

\maketitle

\section{Introduction}\label{sec:1}
Active galactic nuclei (AGNs) are powered by accretion onto supermassive black holes (SMBHs).
Recent observational and theoretical studies suggest that most AGN have multiple absorbers. Broadly speaking, three  types of absorbers have been recognized.
The first one is the {\it warm absorbers} (WAs). WAs are detected as absorption lines and edges in the soft X-ray band of $\sim$65\% of nearby AGNs, especially in the radio-quiet Seyfert 1 galaxies  (e.g., \citealp{McKernan07}, \citealp{Laha14}).
These absorbers are known to be mildly ionized and blue-shifted with a velocity of $\lesssim2,000~ {\rm km/s}$.

The second one is the {\it ultrafast outflow} (UFO). Seyfert galaxies often have blue-shifted, highly ionized absorption lines in their 30--40 \% X-ray energy spectra  \citep{Tombesi10a}. The UFOs are thought to be outflowing winds from the accretion disk at very high velocities of 0.1--0.3 c. Since the UFOs have a larger solid angle than the relativistic jets (e.g., \citealp{Nardini15}, \citealp{Hagino15}), they can supply gas to a wider area of the host galaxy.  UFO contribution to the co-evolution of the supermassive black holes (SMBHs) and the galaxies may be comparable to or even greater than that of the jets (e.g., \citealp{King10}, \citealp{King15}).

The third one is the {\it partial covering absorbers}, which partially absorb the X-ray continuum  in the line of sight. This is also called the ``obscurer'', which causes simultaneous soft X-ray and UV absorption troughs  in  NGC~5548  \citep{Kaastra14}.

\iras is a narrow-line Seyfert 1 (NLS1) galaxy with a SMBH of about $10^{6-7}$~M$_{\odot}$ (z$=$0.0658).
Using 1.5~Ms deep observations of \iras made by \xmm in 2016, \cite{Parker17} discovered UFO absorption lines. They claimed that the equivalent width of the UFO absorption lines and the X-ray luminosity are anti-correlated, while  the line-of-sight velocity of the UFO is correlated with the luminosity. They  argue that the disk wind is driven by the radiation pressure, so  that the absorption lines were not detected when the wind was fully ionized due to intense X-ray radiations.

Two-dimensional radiation-magnetohydrodynamic (R-MHD) simulations of supercritical accretion flows show that the radiation pressure generates a disk wind, and that the wind becomes clumpy due to Rayleigh-Taylor instability at a large distance where the radiation pressure is dominant over the gravitational potential \citep{Takeuchi13}.  In addition to the Rayleigh-Taylor instability, contribution of the radiation hydrodynamic instability in  the clump formation is suggested in  three-dimensional R-MHD simulations   \citep[e.g.,][]{Kobayashi18}.

These inner-wind and outer clumps may be actually observed as the ultrafast outflows (UFOs) and the clumpy absorbers, respectively (``hot inner and clumpy outer wind model''; \citealp{Mizumoto19}).
The clumpy partial absorbers are represented by four observable parameters; partial covering fraction (CF), ionization parameter $\xi$, hydrogen column density \nh, and the line-of-sight velocity \vout.
However, in the standard X-ray spectral analysis,  parameters of the partial absorption by clumpy absorbers and other components such as WAs and soft-excess are often degenerate in the soft X-ray band. 

In the present  study, we adopt  a new data analysis method which we may  call  ``spectral-ratio model fitting'' to the long campaign data of \iras.
The objective is to disentangle   spectral parameter degeneracy amid the complex spectral variations, and to constrain   parameters of the partial absorbers.

\section{Observations and Methods}\label{sec:2}
\subsection{Observations and Data reduction}
% archival data + table-------------------------------------------------------------
We used the 1.5~Ms data obtained by \xmm EPIC-pn (\citealp{Jansen01a}, \citealp{Struder01a}) in the summer of 2016 (Table~\ref{tab:obs}).
The EPIC-MOS data were not used because its effective area is much smaller  than that of the pn detector in the soft X-ray band.
The pn data were reduced with SAS (version 19.1.0) to obtain the filtered event files.
The unscheduled observations  (See Table~\ref{tab:obs}) were also used for the analysis to improve the statistics\footnote{In case of there were interruptions, for example, due to high radiation levels, the exposures taken after the interruptions are called unscheduled.}.
Good time intervals were calculated by removing the periods dominated by the  flaring particle background when the  {\tt PATTERN==0} count rate in 10--12~keV is larger than 0.4~counts~s$^{-1}$.
We used circular regions of 250~physical unit ($12.5''$) radius and 1600 physical unit ($75''$) radius from the same CCD for extracting the source events and the background events, respectively. These region sizes are determined following to  \cite{Chartas18} to maximize the signal-to-noise ratio.
Also, the background region is carefully selected, avoiding chip edges and the region where the Cu background is high.
Following the ``SAS Data Analysis Threads''\footnote
{\samepage{https://www.cosmos.esa.int/web/xmm-newton/sas-thread-epic-filterbackground\\https://www.cosmos.esa.int/web/xmm-newton/sas-thread-pn-spectrum\\https://www.cosmos.esa.int/web/xmm-newton/sas-thread-rgs}}, we extracted  light curves, spectra, and responses in the  0.3--10.0~keV band applying the  good time intervals and the source/background regions.

\begin{table}
	\centering
	\caption{Observation logs with \xmm (pn)}
	\label{tab:obs}
	\scalebox{1.0}{
	\begin{threeparttable}
\begin{tabular}{lccc} \hline 
Start Date & Obs ID & Net Exp. (s)\tnote{a} \\ \hline 
2016 Jul 08 & 0780560101\tnote{b} & 53,333 \\ 
2016 Jul 10 & 0780561301 & 122,231 \\
2016 Jul 12 & 0780561401\tnote{b} & 113,522 \\
2016 Jul 20 & 0780561501 & 119,582 \\
2016 Jul 22 & 0780561601 & 118,232 \\
2016 Jul 24 & 0780561701 & 117,300 \\ 
2016 Jul 26 & 0792180101 & 122,702 \\
2016 Jul 30 & 0792180201 & 119,916 \\
2016 Aug 01 & 0792180301\tnote{b} & 108,568 \\
2016 Aug 03 & 0792180401 & 108,038 \\
2016 Aug 07 & 0792180501 & 111,833 \\
2016 Aug 09 & 0792180601 & 115,994 \\
\hline\end{tabular}
	\begin{tablenotes}
	\item[a] Sum of all the good time intervals.
	\item[b] Unscheduled observations are included in these observation IDs. We combined the scheduled data and the unscheduled data.
	\end{tablenotes}
	\end{threeparttable}
	}
\end{table}

\begin{figure}
\centerline{\includegraphics[width=1.0\columnwidth]{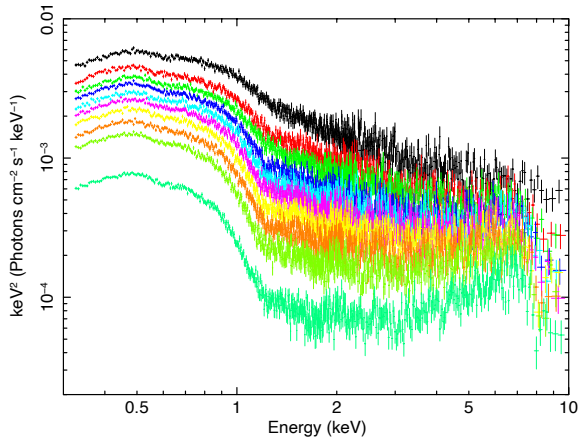}}
	    \caption{The 10  intensity-sliced spectra obtained by \xmm pn in 0.3--10.0~keV. The $\nu F_\nu$ plot unfolded by the power-law with the index of 2 is shown.}
	    \label{fig:sliced}
\end{figure}

% intensity-sliced-------------------------------------------------------------
Since the X-ray spectral features (hardness ratio and absorption line depth etc.) are variable according to the luminosity, we created intensity-sliced spectra following  \cite{Parker17} and \cite{Pinto18}.
First, all the 0.3--10.0~keV events from all the observation IDs are binned every 1~ks, and these time-bins are  sorted in the order of the counts per bin.
The total number of events being  X, we defined 10 groups in order, so that   each group  has  0.1 X events in total.
In this manner, 10 intensity-sliced spectra were created (Table~\ref{tab:sliced}).
The sliced spectra were grouped with 8 and 16
spectral-bins in 0.3--2.0~keV and 2.0--10.0~keV, respectively.
The intensity-sliced spectra are shown in Figure~\ref{fig:sliced}.

\begin{table}
	\centering
	\caption{Count rate threshold of intensity-sliced spectra (0.3--10.0~keV)}
	\label{tab:sliced}
	\scalebox{1.0}{
	\begin{tabular}{lcc}\hline Division & \begin{tabular}{c}Minimum \\(counts s$^{-1}$)\end{tabular}  & \begin{tabular}{c}Maximum \\(counts s$^{-1}$)\end{tabular} \\ \hline 
A & 7.55 & 12.51\\
B & 6.09 & 7.55\\
C & 5.21 & 6.09\\
D & 4.46 & 5.21\\
E & 3.92 & 4.46\\
F & 3.34 & 3.92 \\
G & 2.74 & 3.34\\
H & 2.23 & 2.74 \\
I & 1.63 & 2.23 \\
J & 0.29 & 1.63 \\\hline
    \end{tabular}
    }
\end{table}

\subsection{Spectral components of IRAS~13224$-$3809 and the spectral-ratio analysis}
The following depiction is adopted in the  X-ray spectral model for  this  study.
The continuum consists of the power-law (PL)  and the soft-excess  that contribute mainly above and below $\sim$1~keV, respectively.
The spectral shape of the continuum does not change significantly, while the continuum intensity and/or the PL index may vary.

The continuum is absorbed by several gases such as WA, UFO, and clumpy absorbers.
UFOs are highly ionized and mainly absorb the hard X-rays above $\sim$6~keV.
WAs do not change their geometry or ionization structure significantly in a timescale of a few weeks, while the partial covering fraction of the clumpy absorbers  significantly changes  on shorter timescales
and significantly alters the spectral shape below $\sim$5~keV (e.g., \citealp{DiGesu15}, \citealp{Midooka22}).

Therefore, taking the spectral-ratio below 5~keV  enables us to focus on the variability of the clumpy absorbers, 
canceling out the less time-variable spectral components.
A significant merit of this  method is that  it is not necessary to  determine the spectral shape of the continuum precisely. 

\begin{figure*}
\centerline{\includegraphics[width=2.1\columnwidth]{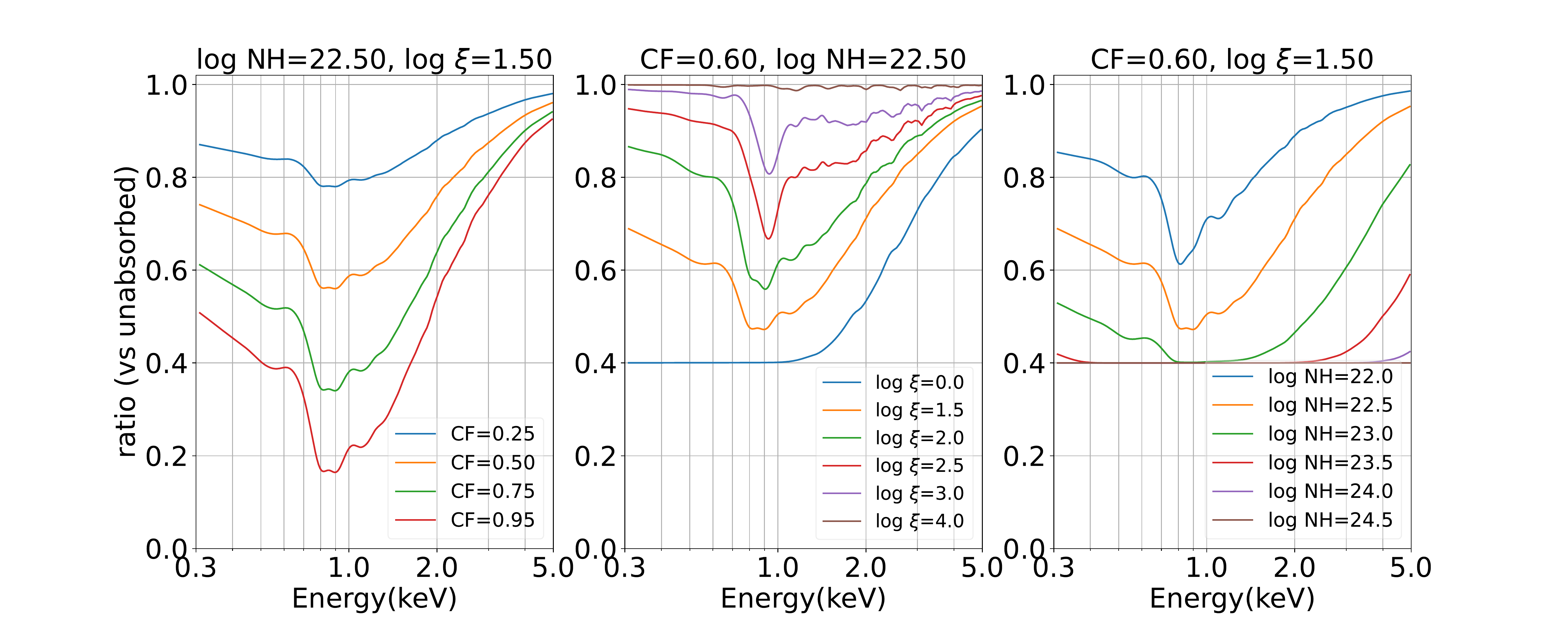}}
	    \caption{Model spectral-ratios, where each  of CF, \nh, and $\xi$ (from left to right) is varied at  several different values while the other  two parameters are fixed.  The absorber velocity is null. 
	    }
	    \label{fig:ratio_model}
\end{figure*}

\subsection{Method of the spectral-ratio analysis} 
Model spectral-ratios are created in the 0.3--5.0~keV band  using  XSPEC (version 12.12.0).
Our aim is to extract the spectral change due to the clumpy absorber.
Therefore, we consider a simple model spectrum (\pl * \zxipcf) in which the continuum \pl is covered by an ionized partial absorber.
Since the response of any intensity-sliced spectra had only variations of less than 1\%, model spectra are extracted using the response of the brightest spectrum A (see Table~\ref{tab:sliced}). 
A table model of the spectral-ratio is created by taking the ratio of the absorbed model spectra to the unabsorbed one with the same continuum parameters.
This operation is repeated for the number of combinations of the parameter grids (Table~\ref{tab:grid}).

Figure~\ref{fig:ratio_model} shows examples of the model spectral-ratios when  each of CF, \nh, and $\xi$ is varied at  several different values  while the  two other parameters  are fixed (the velocity of the absorber is null).
When the absorber is mildly ionized ($\log\xi\sim1.5-3.0$), characteristic dip structures are seen in 0.8--1.0~keV. 
They are mainly created by Fe-L and Fe-M Unresolved Transition Array (UTA), and its depth, width, and shape are dependent on the ionization parameter and the column density. 
Ion population changes as the ionization increases, and the dip energy increases up to $\sim$0.95~keV
It is noteworthy that the dip energy never goes  beyond  $\sim$1 keV, no matter how much the gas is ionized. If  observed spectral-ratios show dips at above 1 keV, that cannot be explained by the ionization, and
strongly suggests that the clumpy absorber is blueshifted.

\begin{figure}
\centerline{\includegraphics[width=1.1\columnwidth]{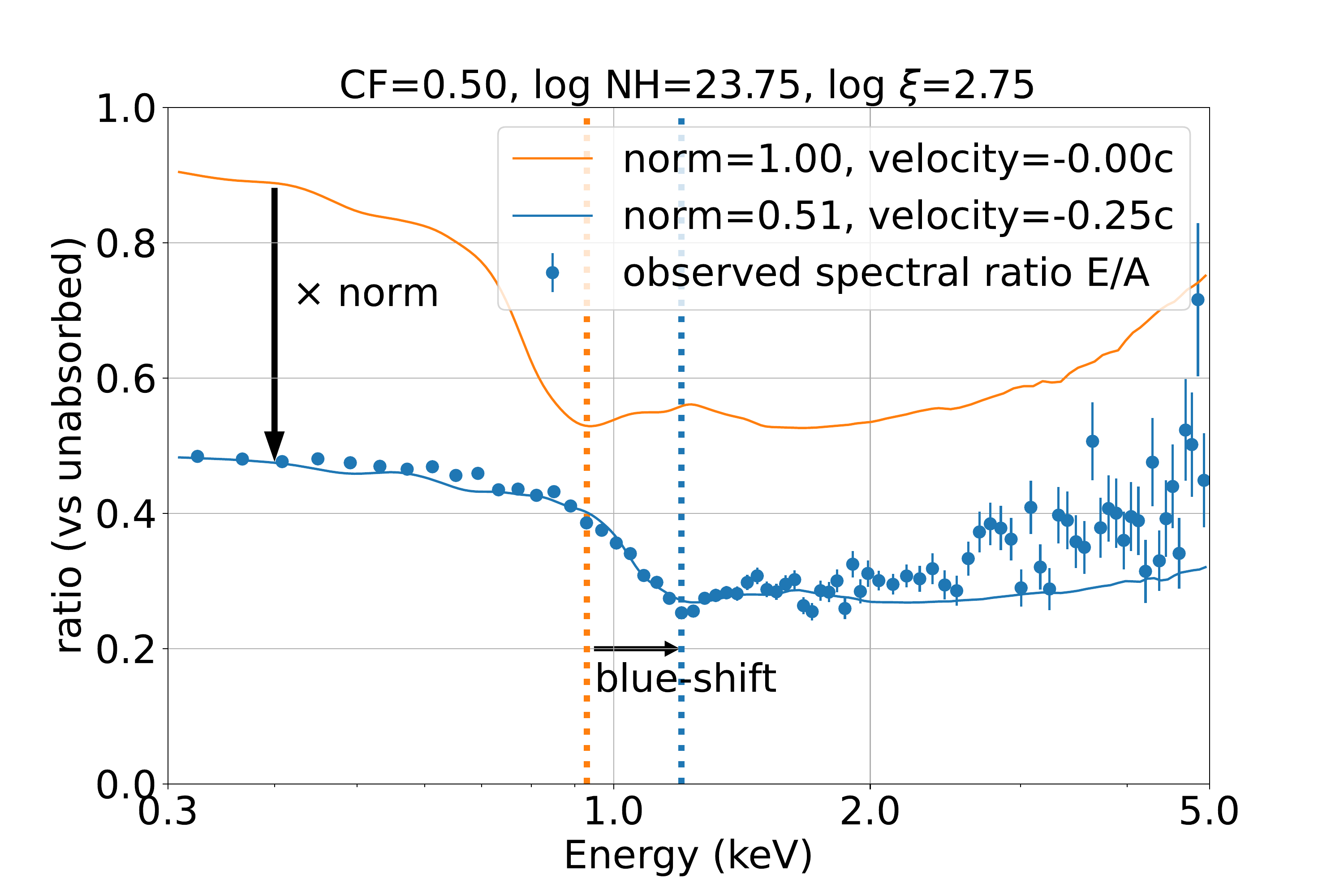}}
	    \caption{Blue dots show the observed spectral-ratio E/A. The orange line represents a model ratio between the absorbed and unabsorbed spectra, where the absorbed spectrum has  CF $= 0.5$, log \nh $= 23.75$, and log $\xi = 2.75$. We find that the dip energy of the model spectral-ratio needs to be blue-shifted. After varying  the norm and velocity, the orange line moves to the blue one, which fits to the observed ratio.
}
	    \label{fig:ratio_EoverA}
\end{figure}

Now we see the observed ratio spectrum.
The ratio of  the fifth brightest spectrum (E)
to the brightest spectrum (A) is shown in blue dots  in Figure~\ref{fig:ratio_EoverA}, where the dip structure is  seen around 1.1--1.3~keV.
We are going to compare the observed spectral-ratio with models,
assuming that the spectrum A is unabsorbed (CF $= 0$).  
We consider an X-ray spectrum with  CF $= 0.5$, log \nh $= 23.75$, log $\xi = 2.75$,
whose  ratio to the unabsorbed  spectrum is the orange curve in Figure~\ref{fig:ratio_EoverA}.
In this spectral-ratio model, the dip structure energy  does not match
with the observation (see Figure~\ref{fig:ratio_model}). %even if the flux variation of the continuum is modeled as ``norm'' and multiplied by a constant between 0 and 1.
Therefore, we introduce the absorber velocity ($-0.25$c)  to blue-shift the dip energy, and also the normalization is made 0.51 times  (blue line in Figure~\ref{fig:ratio_EoverA}).  We see that the observed dip feature is 
explained well by the model spectral-ratio.

\begin{figure}
\centerline{\includegraphics[width=1.1\columnwidth]{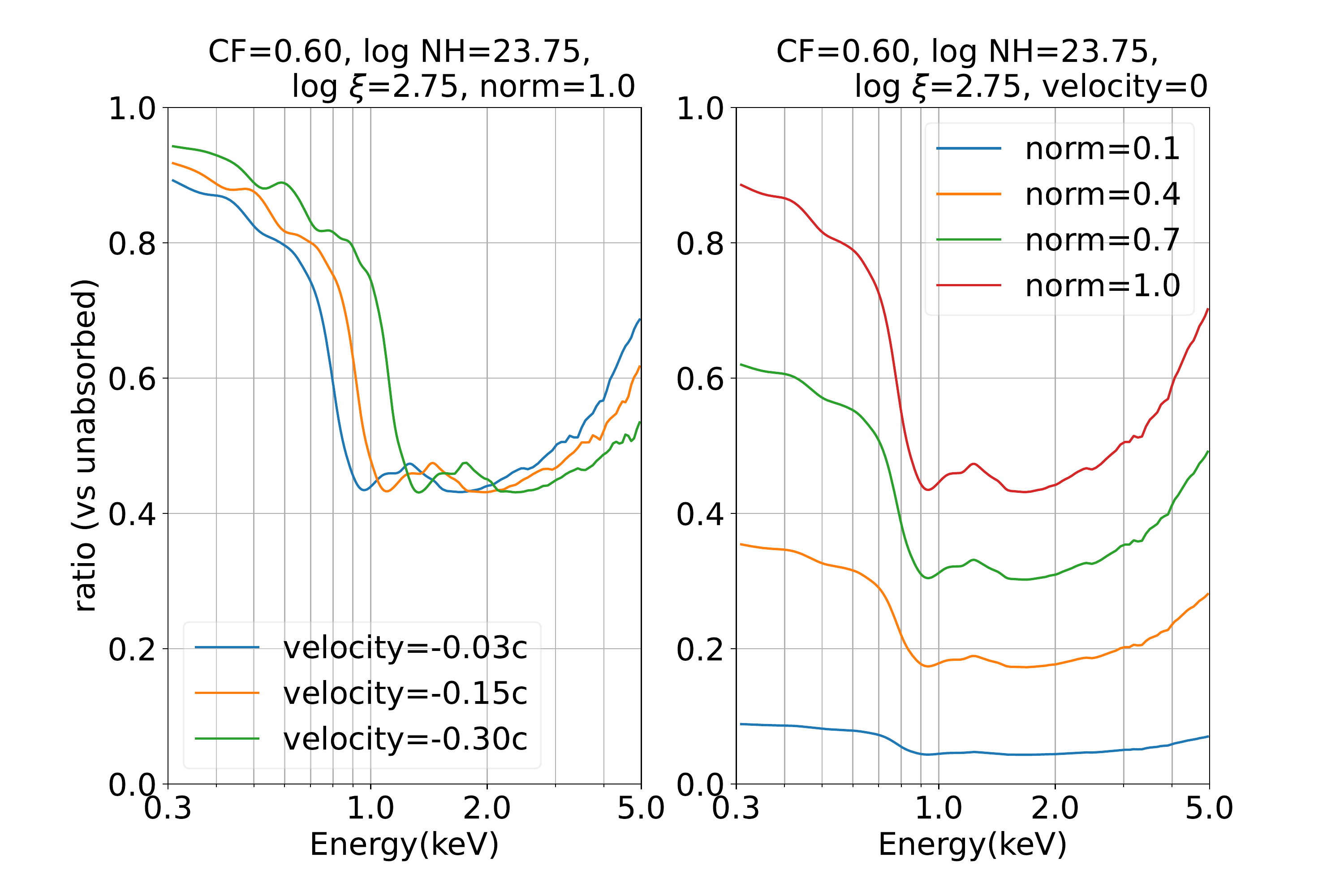}}
	    \caption{Model spectral-ratios, where the velocity (left) or the norm (right) of the absorbed spectrum is  varied at  several different values, while the other parameters are fixed. }
	    \label{fig:ratio_model2}
\end{figure}

Variation of the model spectral-ratios  when the velocity or the norm is  varied
is  shown in Figure~\ref{fig:ratio_model2}.
 The energy shift of the dip structure is apparent with the velocity variation,
so  the absorber velocity is expected to be constrained.

\section{Data Analysis and Results}\label{sec:3}
\subsection{Spectral-ratio fitting in the 0.3--5~keV band} %----------------------------------------------------------------
We made 9  spectral-ratios from the 10 observed intensity-sliced spectra.
Model spectral-ratios were created using the parameter range in Table~\ref{tab:grid}.
Least-square fitting was performed for the 9 observed spectral-ratios, where models are  linearly interpolated between the  adjacent grid points.

\begin{table}
	\centering
	\caption{Parameter ranges for the spectral-ratio model}
	\label{tab:grid}
\begin{tabular}{lcc} \hline 
Component & Minimum & Maximum  \\ \hline 
CF & 0 & 1  \\ 
velocity [c] & $-$0.3 & 0  \\ 
log $\xi$ & 0 & 3  \\ 
\nh & 10$^{22}$ & 10$^{25}$  \\ 
\hline\end{tabular}
\end{table}

\begin{figure*}
\centerline{\includegraphics[width=2.0\columnwidth]{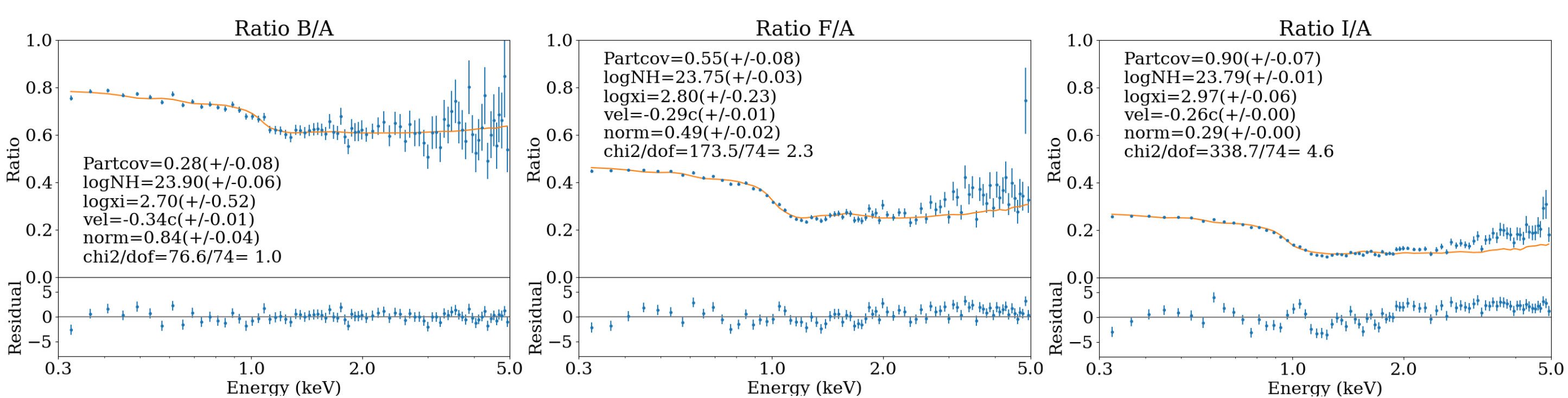}}
	    \caption{Examples of the spectral-ratio model fitting  (B/A, F/A, and I/A). Fitting of the dimmer ones   shows more  residuals above $\sim$3 keV, which may suggest changes in the PL index.}
	    \label{fig:specfit}
\end{figure*}

Figure \ref{fig:specfit} shows examples of the spectral-ratio model fitting, where the lower panel shows residuals normalized by the statistical errors, (data-model)/error.
The free parameters are CF, $\xi$, $N_{\rm H}$, $v$, and norm (which explains the level of the continuum flux).
They are constrained well.
While the \nh and $\xi$ were not significantly variable, the other CF, $v$, and norm vary. 
The largest variation is seen in CF, where the partial covering fraction of the clumpy absorbers varies from 30\%  to 90\%. This produces a characteristic soft X-ray band variation.
The intrinsic flux also changes. The flux at the faintest spectrum is reduced to 30\% of that at the brightest one.
This indicates that the X-ray spectral variations result from a combination of intrinsic luminosity and partial absorber variations.
It is noteworthy that the velocity of the partial absorber is very fast, $\sim0.3c$, and that the velocity increases as the flux gets brighter. This velocity is roughly consistent with that of the UFO. Therefore, it is inferred that clumpy absorbers and UFOs are produced by the same physical mechanisms.

For the fainter spectra at >2~keV, the observed spectral-ratios tend to be higher than the models (middle and right panels in Figure \ref{fig:specfit}), which is mainly due to the current assumption that  the intrinsic spectral shape does not vary.  If the spectra get ``softer when brighter'', as is the case in many AGNs, these residuals could be explained. This topic  will be closely investigated  in our  subsequent paper.

\section{Conclusion}\label{sec:5}
A novel  ``spectral-ratio model fitting'' was developed to resolve  the degeneracy of the soft X-ray spectral components. 
Taking the spectral ratios of the intensity-sliced spectra below 5~keV makes the spectral variations due to partial absorbers conspicuous,  by canceling out the less time-variable continuum and the warm absorber components.
As a result, we succeeded in  constraining the clump outflow velocity, which was  comparable to that of the UFOs. We also found that these velocities get higher with increasing  the X-ray luminosity. 

\section*{Acknowledgments} %----------------------------------------------------------------
This research has made use of data and software provided by the High Energy Astrophysics Science Archive Research Center (HEASARC), which is a service of the Astrophysics Science Division at NASA/GSFC and the High Energy Astrophysics Division of the Smithsonian Astrophysical Observatory. This study was based on observations obtained with \xmm, ESA science missions with instruments and contributions directly funded by ESA Member States and NASA. This research was supported by JSPS Grant-in-Aid for JSPS Research Fellow Grant Number JP20J20809 (T.M.), JSPS KAKENHI Grant Number JP21K13958 (M.M). M.M. acknowledges support from the Hakubi project at Kyoto University.

\bibliography{main}

\end{document}